\documentclass[pre,twocolumn,floatfix,aps,showpacs]{revtex4}

\usepackage{graphicx}
\usepackage{amsmath}
\usepackage{amsfonts}
\usepackage{amssymb}
\usepackage{epsf}
\usepackage{bm}

\begin{document}

\title{Controlling Chaotic Transport on Periodic Surfaces}
\author{R. Chac\'{o}n,$^{1}$ and A.M. Lacasta$^{2}$}
\affiliation{$^{1}$Departamento de F\'{\i}sica Aplicada, Escuela de Ingenier\'{\i}as
Industriales, Universidad de Extremadura, Apartado Postal 382, E-06071
Badajoz, Spain, EU\\
$^{2}$Departament de F\'{\i}sica Aplicada, Universitat Polit\`{e}cnica de
Catalunya, Avinguda Doctor Mara\~{n}\'{o}n 44, E-08028 Barcelona, Spain, EU}
\date{\today}

\begin{abstract}
We uncover and characterize different chaotic transport scenarios on perfect
periodic surfaces by controlling the chaotic dynamics of particles subjected
to periodic external forces in the absence of a ratchet effect. After
identifying relevant {\it symmetries} of chaotic solutions, analytical estimates
in parameter space for the occurrence of different transport scenarios are
provided and confirmed by numerical simulations. These scenarios are highly
sensitive to variations of the system's asymmetry parameters, including the
eccentricity of the periodic surface and the direction of dc and ac forces,
which could be useful for particle sorting purposes in those cases where
chaos is unavoidable.
\end{abstract}

\maketitle

\textbf{\ }\textit{Introduction.}$-$Controlling the transport of particles
on periodic potential energy surfaces is an old and ubiquitous problem
appearing in different fields such as physics, chemistry, and biology \cite{1}.
Specific examples include colloidal transport in arrays of optical tweezers
\cite{2}, flux creep through type-II superconductors \cite{3}, and Bose-Einstein
condensates with periodic pinning sites \cite{4}, among many other. Previous
theoretical analysis of the motion of particles on surfaces \cite{5,6,7,8,9,10}
considered mesoscopic models owing to the great complexity of the different
transport scenarios. While non-chaotic regimes have been widely studied in
the context of noisy overdamped models \cite{11} and the chaotic regime has been
mainly considered when directed transport is induced by symmetry breaking
\cite{9,12}, to the best of our knowledge, the fundamental case of deterministic
chaotic transport in the absence of a ratchet effect has not been considered
in detail as yet. The study of such a chaotic transport on simple periodic
surfaces could indeed shed some light on diverse chaotic phenomena of great
complexity appearing for example in magnetotransport on antidot lattices
\cite{13}.

\textit{Model.}$-$In this Letter, we consider the classical dynamics of a
dissipative particle moving on a standard separable periodic potential, with
an external force having both dc and ac components, and neglecting thermal
effects: $m\overset{..}{x}+\partial V/\partial x=-\mu \overset{.}{x}%
+f_{0}\cos \theta +f_{1x}\cos \left( \omega _{x}t\right) $, $m\overset{..}{y}%
+\partial V/\partial y=-\mu \overset{.}{y}+f_{0}\sin \theta +f_{1y}\cos
\left( \omega _{y}t\right) $, where an overdot denotes a derivative with
respect to $t$, $\theta $ describes the direction of the dc force $\mathbf{f}%
_{0}$, $\mu $ is the phenomenological coefficient of friction, and $%
V(x,y)=V_{0}\left[ \cos \left( 2\pi x/\lambda _{x}\right) +\cos \left( 2\pi
y/\lambda _{y}\right) \right] /2$ is the potential with $\lambda
_{x},\lambda _{y}$ being the characteristic length scales. A main purpose of
the present work is a theoretical characterization of the different chaotic
transport (CT) scenarios by providing analytical estimates of the threshold
conditions in parameter space by using Melnikov analysis (MA). For the sake
of a dimensionless description, we put the equations of motion into the form%
\begin{eqnarray}
\overset{..}{r}_{x}+\sin r_{x} &=&-\gamma \overset{.}{r}_{x}+F_{0x}\cos
\theta +F_{1x}\cos \left( \Omega _{x}\tau \right) ,   \\
\overset{..}{r}_{y}+\frac{\sin r_{y}}{a^{2}} &=&-\gamma \overset{.}{r}_{y}+%
\frac{F_{0x}}{a}\sin \theta +\frac{F_{1x}b}{a}\cos \left( c\Omega _{x}\tau
\right) ,  
\end{eqnarray}%
where all variables and parameters are dimensionless, an overdot denotes a
derivative with respect to $\tau \equiv \pi t\left( 2V_{0}/m\right)
^{1/2}/\lambda _{x}$, $r_{x}\equiv 2\pi x/\lambda _{x}\pm \pi $, $%
r_{y}\equiv 2\pi y/\lambda _{y}\pm \pi $, $\gamma \equiv \mu \lambda
_{x}\left( 2mV_{0}\right) ^{-1/2}/\pi $, $F_{0x}\equiv f_{0}\lambda
_{x}/\left( \pi V_{0}\right) $, $F_{1x}\equiv \lambda _{x}f_{1x}/\left( \pi
V_{0}\right) $, $\Omega _{x}\equiv \omega _{x}\lambda _{x}\left[ m/\left(
2V_{0}\right) \right] ^{1/2}/\pi $, $a\equiv \lambda _{y}/\lambda _{x}$, $%
b\equiv f_{1y}/f_{1x}$, and $c\equiv \omega _{y}/\omega _{x}$. It is also
assumed that the system [Eqs. (1)-(2)] satisfies the MA requirements, i.e.,
the dissipation and forcing terms are small-amplitude perturbations of the
underlying conservative pendula $\overset{..}{r}_{x,y}+\sin r_{x,y}=0$ (see
\cite{14,15,16} for general background). Straightforward application of MA to
Eqs.~(1) and (2) yields the Melnikov functions (MFs)%
\begin{eqnarray}
M_{x}^{\pm }\left( \tau _{0}\right) &=& D_{x}^{\pm }\pm 2\pi F_{1x} \rm{sech}%
\left( \frac{\pi \Omega _{x}}{2}\right) \cos \left( \Omega _{x}\tau_{0}\right) ,  \\
M_{y}^{\pm }\left( \tau _{0}\right) &=& D_{y}^{\pm }\pm 2\pi abF_{1x}\rm{sech}%
\left( \frac{\pi a c\Omega _{x}}{2}\right) \cos \left( c\Omega _{x}\tau_{0}\right) ,  
\end{eqnarray}%
respectively, where the positive (negative) sign refers to the top (bottom)
homoclinic orbit of the conservative pendulum, and $D_{x}^{\pm }\equiv \pm
2\pi F_{0x}\cos \theta -8\gamma $, $D_{y}^{\pm }\equiv \pm 2\pi aF_{0x}\sin
\theta -8a\gamma $. Since the MFs (3) and (4) have an infinity of simple
zeros, a main conclusion is that necessary conditions for the onset of
chaotic instabilities are, respectively,%
\begin{eqnarray}
F_{1x} &>&\frac{\min \left\{ \left\vert D_{x}^{+}\right\vert ,\left\vert
D_{x}^{-}\right\vert \right\} }{2\pi }\cosh \left( \frac{\pi \Omega _{x}}{2}%
\right) ,   \\
F_{1x} &>&\frac{\min \left\{ \left\vert D_{y}^{+}\right\vert ,\left\vert
D_{y}^{-}\right\vert \right\} }{2\pi ab}\cosh \left( \frac{\pi ac\Omega _{x}%
}{2}\right) .  
\end{eqnarray}%
Next, one can compare the theoretical predictions and Lyapunov exponent (LE)
calculations \cite{15} with the caveat that one cannot expect too good a
quantitative agreement between the two kinds of results because LE provides
information concerning solely steady chaos, while MM is a perturbative
method generally related to transient chaos \cite{16}. To quantify the sorting
capability associated with the threshold of chaotic transport, we evaluate
the Cartesian components of the velocity, $\left\langle v_{i}\right\rangle
=\lim_{\tau \rightarrow \infty }\left\langle r_{i}(\tau )\right\rangle /\tau 
$ $\left( i=x,y\right) $, where brackets indicate average over initial
conditions, and construct the velocity components parallel and perpendicular
to the external dc force $\mathbf{f}_{0}$, $\left\langle v_{\Vert
}\right\rangle =\left\langle v_{x}\right\rangle \cos \theta +a\left\langle
v_{y}\right\rangle \sin \theta $, and $\left\langle v_{\bot }\right\rangle
=-\left\langle v_{x}\right\rangle \sin \theta +a\left\langle
v_{y}\right\rangle \cos \theta $, respectively. We characterize the
deviation of $\left\langle \mathbf{v}\right\rangle $ from $\mathbf{f}_{0}$
by means of the quantifier 
\begin{equation}
\tan \alpha =\left\langle v_{\bot }\right\rangle /\left\langle v_{\Vert
}\right\rangle ,  
\end{equation}%
where $\alpha $ is the deflection angle \cite{10}. For the sake of clarity, we
shall consider here the case with equal frequencies $\left( c=1\right) $ and
both dc and ac forces acting in the same direction ($f_{1x}\equiv f_{1}\cos
\theta ,f_{1y}\equiv f_{1}\sin \theta $, and hence $b=\tan \theta $). By
defining $F_{1}\equiv \lambda _{x}f_{1}/\left( \pi V_{0}\right) $, one has $%
F_{1x}=F_{1}\cos \theta $ and hence Eqs. (5) and (6) reduce to%
\begin{eqnarray}
F_{1} &>&F_{1,th}^{x}\equiv \frac{\min \left\{ \left\vert
D_{x}^{+}\right\vert ,\left\vert D_{x}^{-}\right\vert \right\} }{2\pi
\left\vert \cos \theta \right\vert }\cosh \left( \frac{\pi \Omega _{x}}{2}%
\right) ,   \\
F_{1} &>&F_{1,th}^{y}\equiv \frac{\min \left\{ \left\vert
D_{y}^{+}\right\vert ,\left\vert D_{y}^{-}\right\vert \right\} }{2\pi
a\left\vert \sin \theta \right\vert }\cosh \left( \frac{\pi a\Omega _{x}}{2}%
\right) ,  
\end{eqnarray}%
respectively, where $F_{1,th}^{x},F_{1,th}^{y}$ are the chaotic threshold
amplitudes.

\textit{Symmetry analysis.}$-$Equations (8) and (9) tell us that the onset
of chaos in both directions strongly depends upon the external force
direction $\theta $, which can thus be used as a \textit{high}-sensitivity
control parameter to suppress and strength CT in one or the another
direction at will. Specifically, one straightforwardly obtains from Eqs. (8)
and (9) that the chaotic threshold amplitudes exhibit (as functions of $%
\theta $) the symmetries:%
\begin{eqnarray}
F_{1,th}^{x}\left( \pi /2\pm \theta \right) &=&F_{1,th}^{x}\left( \pi /2\mp
\theta \right) ,   \\
F_{1,th}^{y}\left( \pi /2\pm \theta \right) &=&F_{1,th}^{y}\left( \pi /2\mp
\theta \right) ,   \\
F_{1,th}^{x}\left( \pi /4\pm \theta \right) &=&F_{1,th}^{y}\left( \pi /4\mp
\theta \right) ,   \\
F_{1,th}^{x}\left( 3\pi /4\pm \theta \right) &=&F_{1,th}^{y}\left( 3\pi
/4\mp \theta \right) .  
\end{eqnarray}%
Now, the following remarks are in order. First, Eqs. (10) and (11) are valid
for any spatial potential $\left( a>0\right) $, while Eqs. (12) and (13) are
solely valid for a symmetric potential $\left( a=1\right) $. Second,
symmetries (12) and (13) imply that different transport regimes are expected
in the $x$- and $y$-directions as the external force direction deviates from
the \textquotedblleft symmetric\textquotedblright\ angles $\pi /4$ and $3\pi
/4$, respectively. Third, in the absence of multistability (i.e., when a
single attractor exists for all initial conditions), symmetries (12) and
(13) also imply $\left\langle v_{x}\right\rangle \left( \pi /4\pm \theta
\right) \simeq \left\langle v_{y}\right\rangle \left( \pi /4\mp \theta
\right) $ and $\left\langle v_{x}\right\rangle \left( 3\pi /4\pm \theta
\right) \simeq \left\langle v_{y}\right\rangle \left( 3\pi /4\mp \theta
\right) $, respectively, and hence $\tan \alpha $ (as a function of $\theta $%
) exhibits the symmetry%
\begin{eqnarray}
\tan \alpha \left( \pi /4+\theta \right) &=&-\tan \alpha \left( \pi
/4-\theta \right) ,   \\
\tan \alpha \left( 3\pi /4+\theta \right) &=&-\tan \alpha \left( 3\pi
/4-\theta \right) ,  
\end{eqnarray}%
i.e., for a symmetric potential, $\tan \alpha $ is an \textit{odd} function
of $\theta $ with respect to the angles $\pi /4$ and $3\pi /4$,
respectively. Note that this is no longer the case for an asymmetric
potential according to the first remark.

\textit{Numerical results.}$-$Extensive numerical simulations confirmed all
the above theoretical predictions. Thus, by varying $\theta $ one can find
different transport regimes (see Fig. 1, top panel): CT in both directions
(as for $\theta =\left\{ 2\pi /9,5\pi /18\right\} $), CT in one direction
while intermittent periodic transport (PT) in the other (as for $\theta
=\left\{ \pi /6,\pi /3\right\} $), PT in both directions (as for $\theta
=\left\{ 7\pi /36,\pi /4,11\pi /36\right\} $), and PT in one direction while
periodic oscillation in the other (as for $\theta =\left\{ 13\pi /36,5\pi
/36\right\} $). Since the onset of chaos also depends upon the particle mass
(through the coefficient of friction, cf. Eqs. (8) and (9)), such an $\theta 
$-dependence can therefore be used to sort different particles according to
their mass. For two kinds of particles with different masses, this means
that one can obtain analytical estimates of the optimal force directions, $%
\theta _{opt}$, from Eqs. (8) and (9) such that one particle exhibits CT
while the other does not, the remaining parameters being held constant.
Numerical experiments confirmed this scenario as is shown in Figs. 1 (medium
panel) and 2. Additionally, the onset of chaos also depends upon the
eccentricity parameter $a$ (Eq. (9)): Decreasing or increasing $a$ from 1
(symmetric potential) means increasing the potential's asymmetry. Thus, the
eccentricity of the periodic potential can also be used as an effective
parameter to control CT on a periodic surface, as in the case of optical
potentials for example \cite{17}. 
\begin{figure}
\begin{center}
\includegraphics[width=0.45\textwidth]{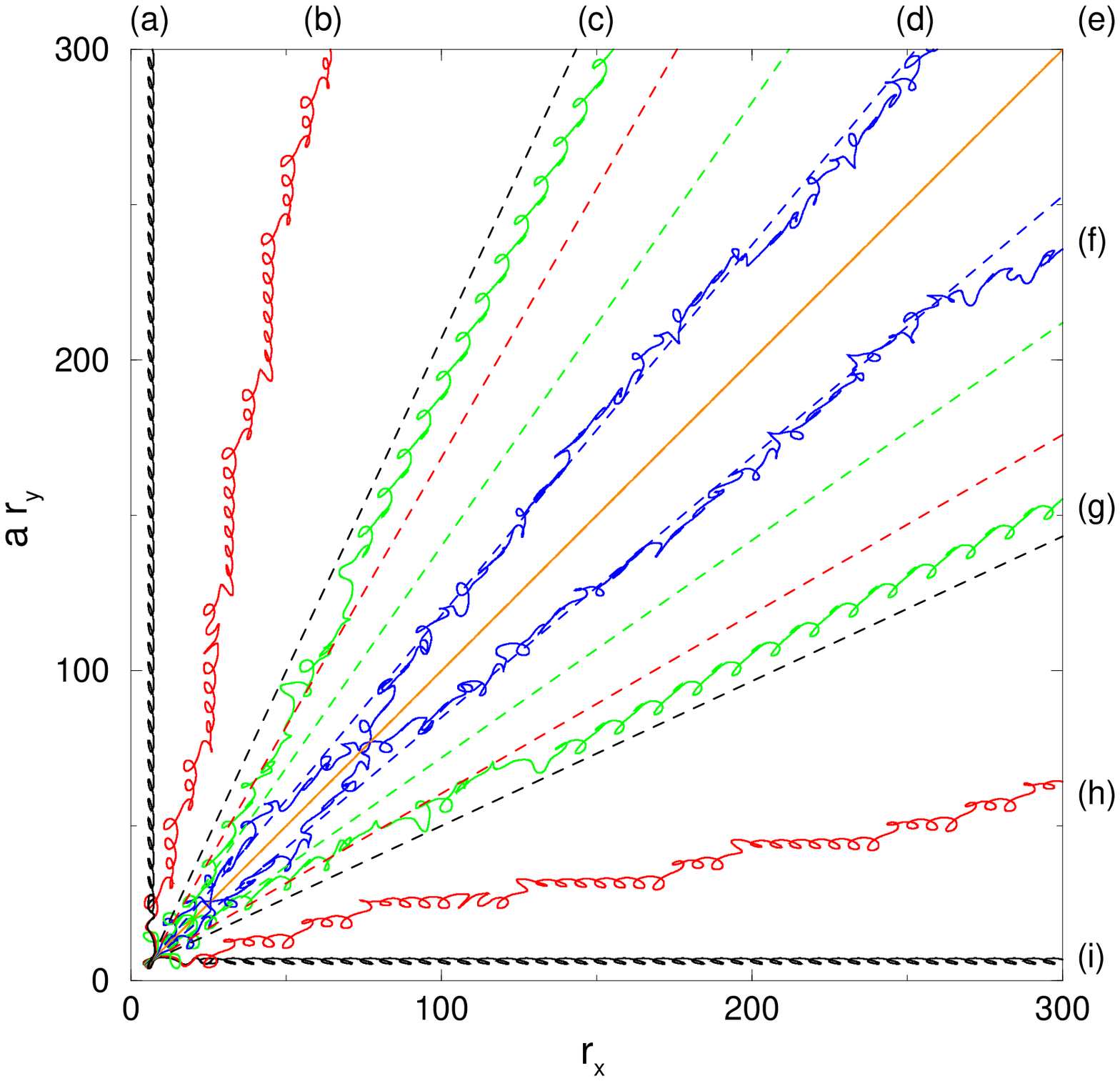}\\
\includegraphics[width=0.45\textwidth]{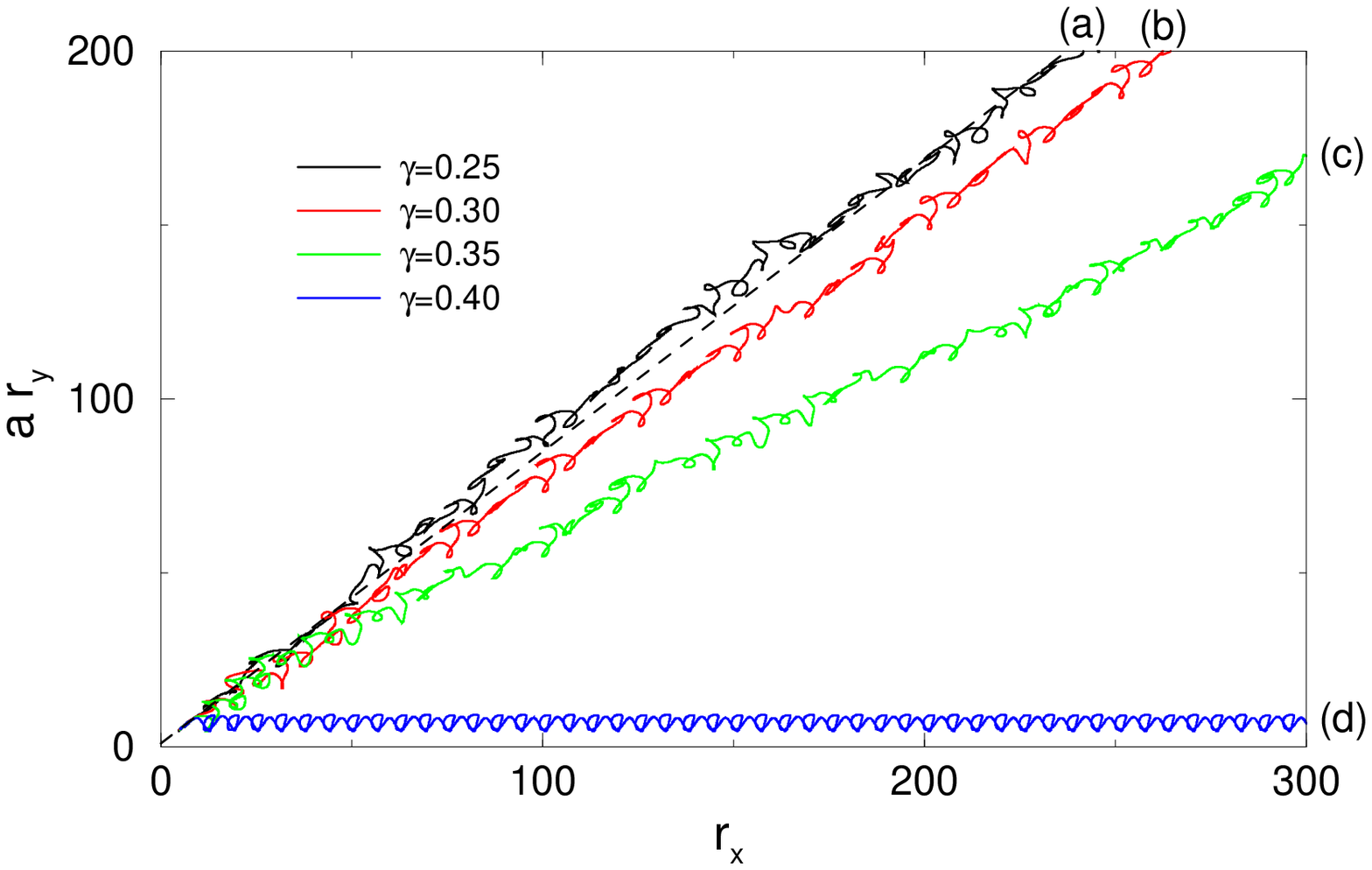}\\
\includegraphics[width=0.45\textwidth]{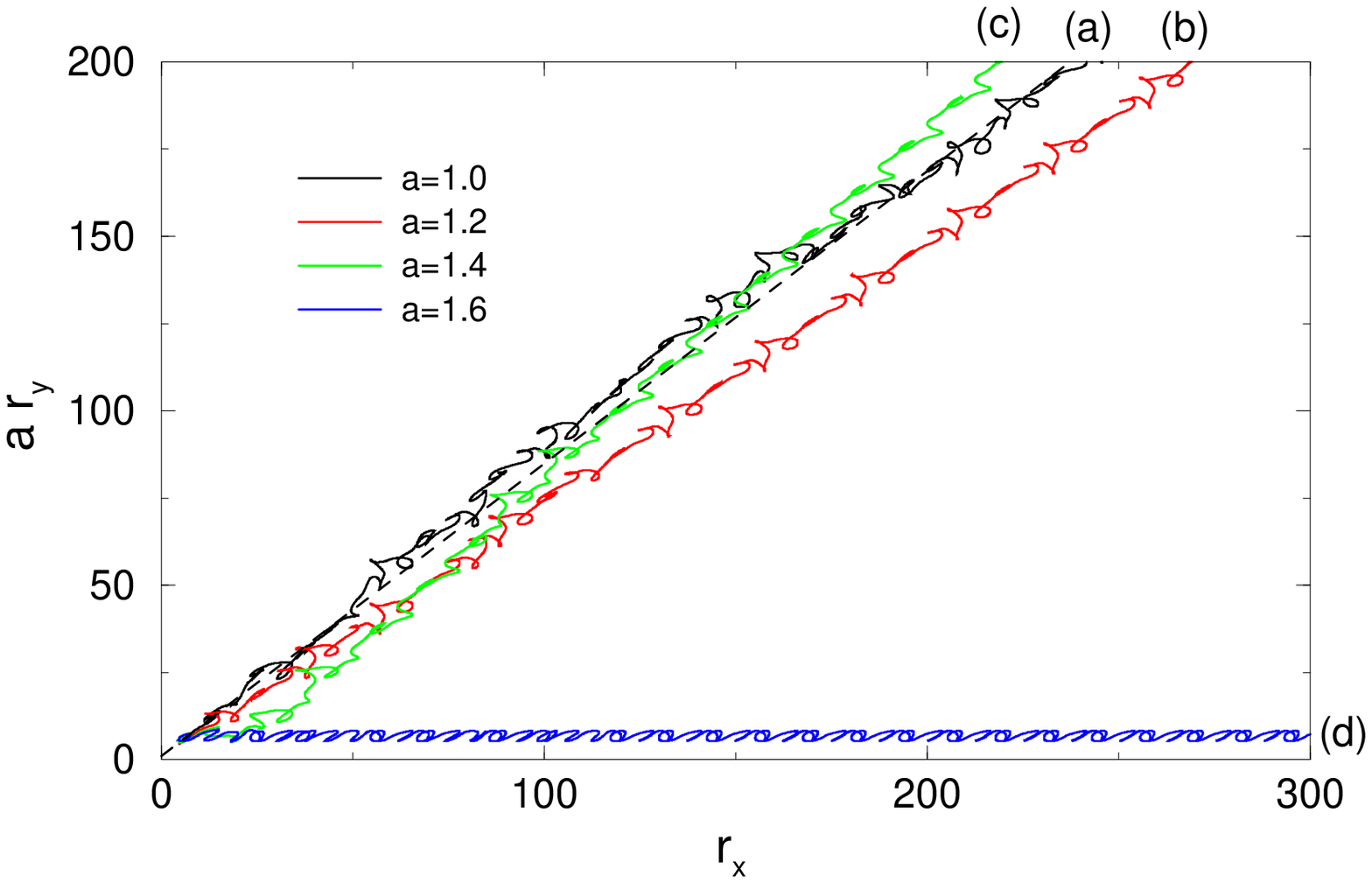}
\end{center}
\caption{
Top panel: Trajectories for a net force applied at different
angles: $\theta =13\pi /36$ (a), $\pi /3$ (b), $11\pi /36$ (c), $5\pi /18$
(d), $\pi /4$ (e), $2\pi /9$ (f), $7\pi /36$ (g), $\pi /6$ (h), and $5\pi
/36 $ (i) for $a=1$ and $\gamma =0.25$. Medium panel: Trajectories for $%
a=1,\theta =2\pi /9$, and four values of the dimensionless coefficient of
friction: $\gamma =0.25$ (a), $0.3$ (b), $0.35$ (c), and $0.4$ (d). Bottom
panel: Trajectories for $\theta =2\pi /9,\gamma =0.25$, and four values of the
eccentricity parameter: $a=1$ (a), $1.2$ (b), $1.4$ (c), and $1.6$ (d).
Other parameters are: $F_{0x}=0.28,F_{1}=1,$ and $\Omega _{x}=0.68$. Dotted
lines indicate the direction of the external force.
}
\label{fig1}
\end{figure}
\begin{figure}
\begin{center}
\includegraphics[width=0.45\textwidth]{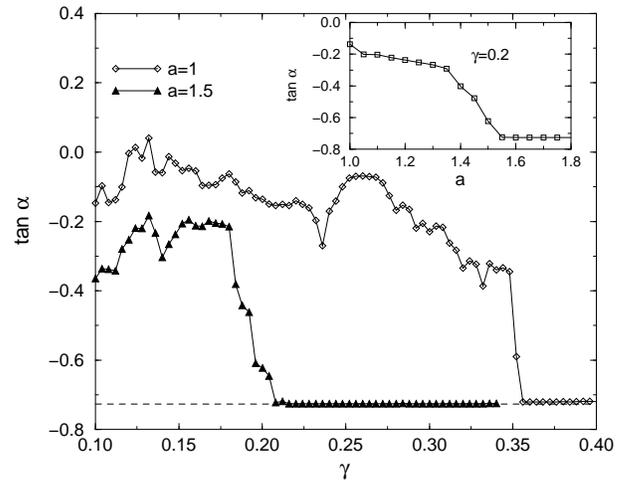}
\end{center}
\caption{
Deflection angle vs coefficient of friction for two values of the
eccentricity parameter: $a=1$ $\left( \Diamond \right) ,1.5$ $\left(
\blacktriangle \right) $. The inset shows the deflection angle vs
eccentricity parameter for $\gamma =0.2$. Other parameters are: $\theta =\pi
/5,F_{0x}=0.28,F_{1}=1,\Omega _{x}=0.68$. The solid lines are solely plotted
to guide the eye.
}
\label{fig2}
\end{figure}
\begin{figure}
\begin{center}
\includegraphics[width=0.45\textwidth]{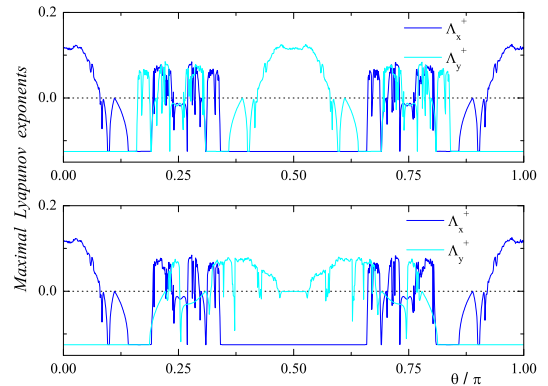}
\end{center}
\caption{
Maximal LEs $\Lambda _{x}^{+},\Lambda _{y}^{+}$ as a function of
the angle $\theta $ for two values of the eccentricity parameter: $a=1$ (top
panel), $1.5$ (bottom panel). Other parameters are: $F_{0x}=0.28,F_{1}=1,%
\gamma =0.25$ and $\Omega _{x}=0.68$.
}
\label{fig3}
\end{figure}
\begin{figure}
\begin{center}
\includegraphics[width=0.45\textwidth]{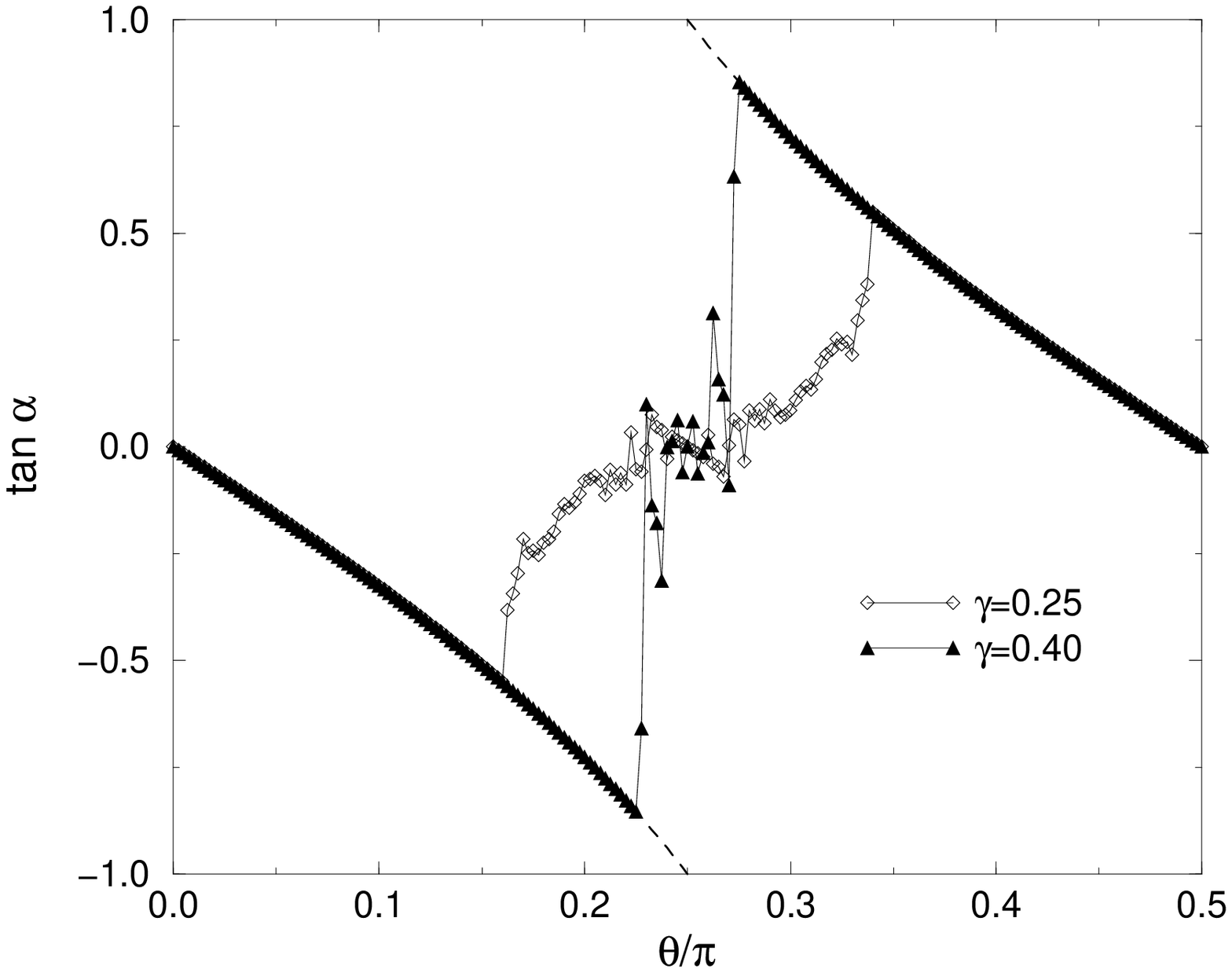}\\
\includegraphics[width=0.45\textwidth]{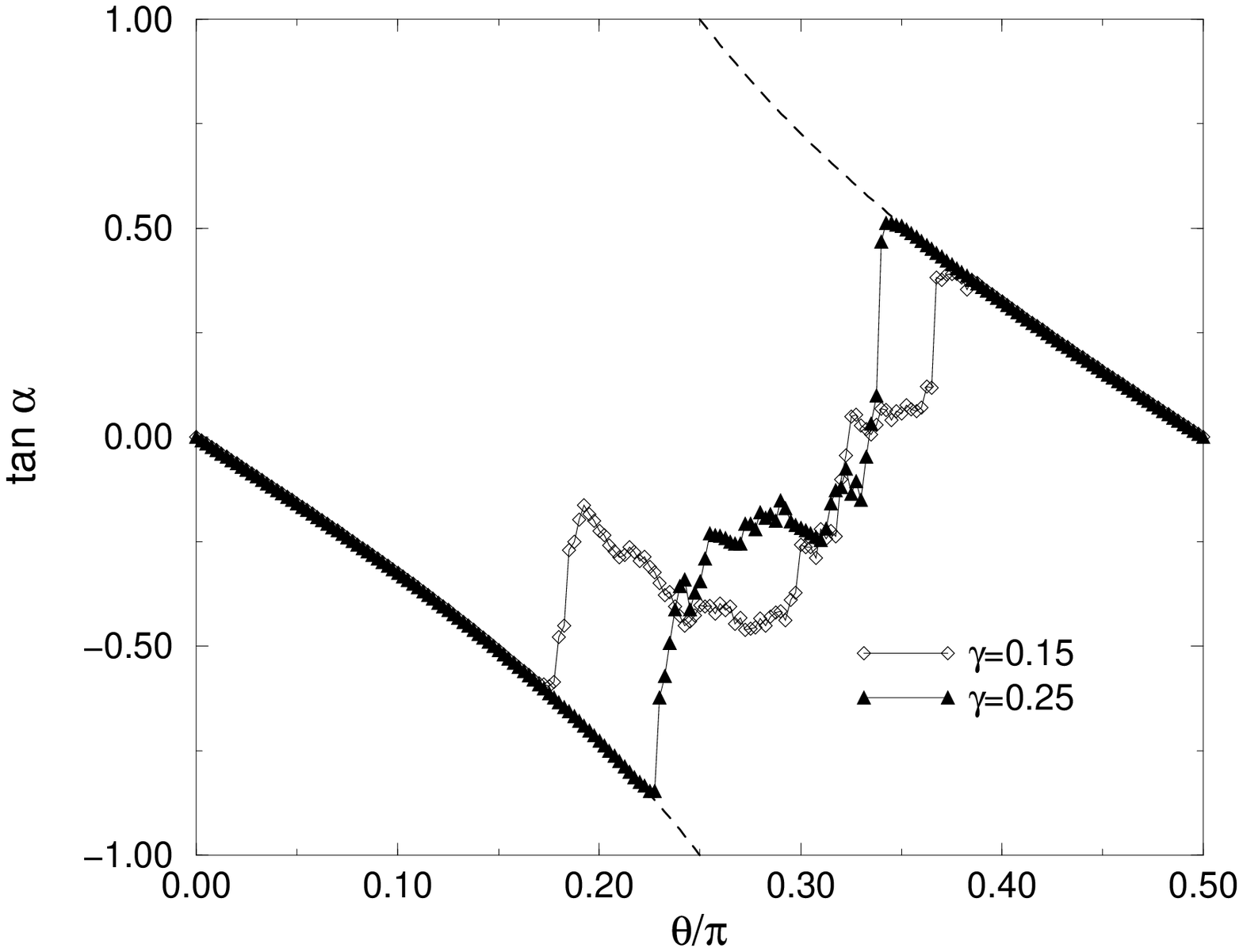}
\end{center}
\caption{
Deflection angle vs external force direction for $%
F_{0x}=0.28,F_{1}=1,\Omega _{x}=0.68$, two values of the eccentricity
parameter: $a=1$ (top panel), $1.5$ (bottom panel), and different values of
the coefficient of friction: $\gamma =0.25\left( \Diamond \right) ,0.4\left(
\blacktriangle \right) $ (top panel) and $\gamma =0.15\left( \Diamond
\right) ,0.25\left( \blacktriangle \right) $ (bottom panel). Also plotted
are the functions $-\tan \theta $ and $\cot \theta $ (dashed lines, see the
text ).
}
\label{fig4}
\end{figure}
Figure 1 (bottom panel) shows an illustrative
example where typical trajectories are plotted for increasing values of $a$
from 1. Starting at a situation where CT occurs in both directions $\left(
a=1\right) $, one finds that increasing the potential's asymmetry $\left(
a>1\right) $ changes the motion to PT in both directions (as for $a=1.2$).
This behaviour changes again to CT in both directions for higher values of $%
a $ (as for $a=1.4$), and finally changes to PT in the $x$-direction while
remain bounded inside a well in the $y$-direction (as for $a=1.6$). Also,
the effectiveness of a fixed external force at sorting heavy particles is
enhanced by breaking the potential symmetry (recall that $\gamma \sim
m^{-1/2}$, see Fig. 2). Figure 3 shows illustrative instances of maximal
LEs, $\Lambda _{x}^{+}$ and $\Lambda _{y}^{+}$, which quantify the chaotic
dynamics in the $x$- and $y$-directions, respectively, versus $\theta $ for
two values of the eccentricity parameter. Remarkably, these diagrams present
relevant symmetries which are coherent with those of the chaotic threshold
amplitudes [Eqs. (10)-(13), respectively]: $\Lambda _{x}^{+}\left( \pi /2\pm
\theta \right) =\Lambda _{x}^{+}\left( \pi /2\mp \theta \right) $, $\Lambda
_{y}^{+}\left( \pi /2\pm \theta \right) =\Lambda _{y}^{+}\left( \pi /2\mp
\theta \right) $, $\Lambda _{x}^{+}\left( \pi /4\pm \theta \right) =\Lambda
_{y}^{+}\left( \pi /4\mp \theta \right) $, $\Lambda _{x}^{+}\left( 3\pi
/4\pm \theta \right) =\Lambda _{y}^{+}\left( 3\pi /4\mp \theta \right) $. It
is worth mentioning that this coherence is far from trivial in the sense
that, to the best of our knowledge, there is no theoretical connection
between MA predictions and LEs for the present system, \textit{thus
indicating the relevance and depth of the chaotic threshold symmetries in
parameter space}. One typically finds how different chaotic and non-chaotic
regimes drastically change over certain $\theta $ ranges as the potential
becomes asymmetric. For instance, PT in both directions at $\theta =\left\{
\pi /4,3\pi /4\right\} $ for a symmetric potential $\left( a=1\right) $
changes to CT in solely one direction for an asymmetric potential $\left(
a=1.5\right) $ (cf. Fig. 3). Finally, numerical simulations confirmed the
accuracy of predictions (14) and (15) as is shown in Fig. 4. Starting with
CT in both directions at $\theta =\pi /4$ for a symmetric potential (Fig. 4,
top panel), one sees that the deflection of particles increases as $\theta $
deviates from $\pi /4$ according to the route described in Fig. 1, top
panel. Maximum deflection occurs at symmetric angles with respect to $\pi
/4,\theta _{\max }^{low},\theta _{\max }^{\sup }\left( \pi /4-\theta _{\max
}^{low}\simeq \theta _{\max }^{\sup }-\pi /4\right) $, where there is PT in
one direction while periodic oscillation in the other. For $\theta \geqslant
\theta _{\max }^{\sup }$ $\left( \theta \leqslant \theta _{\max
}^{low}\right) $, this transport regime remains, i.e., $\left\langle
v_{x}\right\rangle =0$ $\left( \left\langle v_{y}\right\rangle =0\right) $
and hence $\tan \alpha \left( \theta \geqslant \theta _{\max }^{\sup
}\right) =\cot \theta $ $\left( \tan \alpha \left( \theta \leqslant \theta
_{\max }^{low}\right) =-\tan \theta \right) $ (cf. Eq. (7)). For an
asymmetric potential (Fig. 4, bottom panel), the dependence of the
deflection angle on the external force direction essentially presents a
similar scenario to that of the symmetric case, but now $\tan \alpha $ is no
longer an odd function with respect $\pi /4$, as predicted (cf. third
remark).

\textit{Conclusions.}$-$To sum, we have demonstrated theoretically and
numerically through a simple and general system that reliable control of
sorting on periodic surfaces is achieved for chaotic particles by
identifying the relevant symmetries of the chaotic threshold in parameter
space. We uncovered and characterized different sorting scenarios associated
with symmetric and asymmetric spatial potentials, which could motivate
experiments in different contexts such as optical and antidot lattices.
Among the most interesting extensions of this work are the case with the ac
and dc forces having different directions, where preliminary results
indicate the presence of intriguing \textquotedblleft absolute negative
mobility\textquotedblright\ phenomena \cite{18}, as well as the study of the
effect of noise on the present transport scenarios: Even very small amounts
of noise may cause both a transition from a bounded state to a running\
state and a significative modification of the chaotic threshold in parameter
space \cite{19}. Our current work is aimed at exploring these cases.

We thank Katja Lindenberg for useful discussions. This work was partially supported 
by the Ministerio de Ciencia e Innovaci\'{o}n (MCINN, Spain) under projects FIS2008-01383 (R. Ch.) 
and  FIS2009-13360-C03-03 (A.M.L.).


\begin{thebibliography}{99}
\bibitem{1} H. Risken, \textit{The Fokker-Planck Equation} (Springer,
Berlin, 1984), Chap. 11.

\bibitem{2} P. T. Korda, M. B. Taylor, and D. G. Grier, Phys. Rev. Lett. 
\textbf{89}, 128301 (2002).

\bibitem{3} C. Reichhardt and F. Nori, Phys. Rev. Lett. \textbf{82}, 414
(1999).

\bibitem{4} J. W. Reijnders and R. A. Duine, Phys. Rev. Lett. \textbf{93},
060401 (2004).

\bibitem{5} A. W. Ghosh and S. V. Khare, Phys. Rev. Lett. \textbf{84}, 5243
(2000).

\bibitem{6} J. D. Bao and Y. Z. Zhuo, Phys. Lett. A \textbf{239}, 228 (1998).

\bibitem{7} I. Der\'{e}nyi and R. D. Astumian, Phys. Rev. E \textbf{58},
7781 (1998).

\bibitem{8} R. Eichhorn, P. Reimann, and P. H\"{a}nggi, Phys. Rev. Lett. 
\textbf{88}, 190601 (2002).

\bibitem{9} R. Guantes and S. Miret-Art\'{e}s, Phys. Rev. E \textbf{67},
046212 (2003).

\bibitem{10} A. M. Lacasta, J. M. Sancho, A. H. Romero, and K. Lindenberg,
Phys. Rev. Lett. \textbf{94}, 160601 (2005).

\bibitem{11} J. P. Gleeson, J. M. Sancho, A. M. Lacasta, and K. Lindenberg,
Phys. Rev. E \textbf{73}, 041102 (2006).

\bibitem{12} S. Denisov, Y. Zolotaryuk, S. Flach, and O. Yevtushenko, Phys.
Rev. Lett. \textbf{100}, 224102 (2008).

\bibitem{13} M. Khoury, A.M. Lacasta, J.M Sancho, A.H. Romero, K. Lindenberg, 
Phys. Rev. B \textbf{78}, 155433 (2008), and references therein.

\bibitem{14} V. K. Melnikov, Trans. Moscow Math. Soc. \textbf{12}, 1 (1963).

\bibitem{15} See, e.g., A. J. Lichtenberg and M. A. Lieberman, \textit{%
Regular and Chaotic Dynamics} (Springer, New York, 1992), Chaps. 5 and 7.

\bibitem{16} J. Guckenheimer and P. J. Holmes, \textit{Nonlinear
Oscillations, Dynamical Systems, and Bifurcations of Vector Fields}
(Springer, Berlin, 1983).

\bibitem{17} G. Grynberg and C. Robilliard, Phys. Rep. \textbf{355}, 335
(2001).

\bibitem{18} D. Speer, R. Eichhorn, and P. Reimann, Phys. Rev. Lett. \textbf{%
102}, 124101 (2009).

\bibitem{19} P. J. Mart\'{\i}nez and R. Chac\'{o}n, Phys. Rev. Lett. \textbf{%
93}, 237006 (2004).
\end{thebibliography}
\end{document}